# IMPROVEMENT OF PROPERTIES OF SELF-INJECTED AND ACCELERATED ELECTRON BUNCH BY LASER PULSE IN PLASMA, USING PULSE PRECURSOR

Vasyl Maslov[1,2]*, Denys Bondar[2], Iryna Levchuk[1], Ivan Onishchenko[1]

*[1] NSC "Kharkiv Institute of Physics and Technology" NASU*
*Kharkiv, 61108, st. Akademicheskaya 1, Ukraine*
*[2] V.N. Karazin Kharkiv National University*
*4 Svobody Sq., Kharkiv, 61022, Ukraine*
*\*E-mail: vmaslov@kipt.kharkov.ua*

The accelerating gradients in conventional linear accelerators are currently limited to ~100 MV/m. Plasma-based accelerators have the ability to sustain accelerating gradients which are several orders of magnitude greater than that obtained in conventional accelerators. Due to the rapid development of laser technology the laser-plasma-based accelerators are of great interest now. Over the past decade, successful experiments on laser wakefield acceleration of electrons in the plasma have confirmed the relevance of this acceleration. Evidently, the large accelerating gradients in the laser plasma accelerators allow to reduce the size and to cut the cost of accelerators. Another important advantage of the laser-plasma accelerators is that they can produce short electron bunches with high energy. The formation of electron bunches with small energy spread was demonstrated at intense laser–plasma interactions. Electron self-injection in the wake-bubble, generated by an intense laser pulse in underdense plasma, has been studied. With newly available compact laser technology one can produce 100 PW-class laser pulses with a single-cycle duration on the femtosecond timescale. With a fs intense laser one can produce a coherent X-ray pulse. Prof. T. Tajima suggested utilizing these coherent X-rays to drive the acceleration of particles. When such X-rays are injected into a crystal they interact with a metallic-density electron plasma and ideally suit for laser wakefield acceleration. In numerical simulation of authors, performed according to idea of Prof. T.Tajima, on wakefield excitation by a X-ray laser pulse in a metallic-density electron plasma the accelerating gradient of several TV/m has been obtained. It is important to form bunch with small energy spread and small size. The purpose of this paper is to show by the numerical simulation that some precursor-laser-pulse, moved before the main laser pulse, controls properties of the self-injected electron bunch and provides at certain conditions small energy spread and small size of self-injected and accelerated electron bunch.

The accelerating electric fields in metal cavities of conventional accelerators are not more than ~100 MV/m for technical reasons, partly due to breakdown. Plasma can provide much more accelerating electric field which is approximately in $10^3$ times larger than possible in metal cavities of conventional accelerators [1]. As plasma in experiment is inhomogeneous and nonstationary and properties of wakefield changes at increase of its amplitude it is difficult to excite wakefield resonantly by a long sequence of electron bunches [2, 3], to focus sequence [4-8], to prepare sequence from long beam [9-11] and to provide large transformer ratio [13-18]. In [4] the mechanism has been found and in [19-22] investigated of resonant plasma wakefield excitation by a nonresonant sequence of short electron bunches. Owing to successful development of laser physics [1, 23] the method of electron acceleration by short laser pulse in plasma is very perspective now. During last years, promising results have measured in experiments on electron acceleration by wakefield, excited by laser pulse in plasma [23]. The large accelerating wakefield, excited by laser pulse in plasma, provides possibility to decrease the dimensions and to cut the cost of accelerators. Important property of the accelerators, based on the laser-plasma interaction, is that short electron bunches can be self-injected and accelerated to high energy [23]. The electron bunch acceleration with small energy spread was observed at interaction of intense laser pulse with plasma.

The problem at laser wakefield acceleration is that laser pulse quickly destroyed because of its expansion. One way to solve this problem is the use of a capillary as a waveguide for laser pulse. The second way to solve this problem is to transfer its energy to the electron bunches which as drivers accelerate witness. A transition from a laser wakefield accelerator to plasma wakefield accelerator can occur in some cases at laser-plasma interaction [24].

Developed technology of compact intense lasers, which was awarded the Nobel Prize 2018, one can provide 100 PW laser pulses of a very small duration on the fs-timescale. Prof. T.Tajima concluded [25] that using this a femtosecond intense laser one can generate a coherent X-ray pulse. He also proposed [25] using these coherent X-rays to accelerate electrons. This X-ray, at injection into a crystal, can excite wakefield in a metallic-density electron plasma and ideally correspond for laser wakefield acceleration [25].

In [24, 26-28] it has shown that at certain conditions the laser wakefield acceleration is added by a beam-plasma wakefield acceleration.

At the laser acceleration of self-injected electron bunch by plasma wakefield the accelerating gradient about 50 GV/m has been obtained in experiments [23]. In numerical simulation [28], performed according to idea of Prof. T. Tajima [25, 29], on wakefield excitation by a X-ray laser pulse in a metallic-density electron plasma the accelerating gradient of several TV/m has been obtained. To solve the problem of laser pulse expansion, one can use a capillary discharge. The second method [23, 24, 26, 27] of solving this problem is fast energy transfer of driver laser pulse to self-injected electron bunch. This bunch becomes driver-bunch and accelerates next self-injected electron bunch up to larger energy.

It is important to form bunch with small energy spread and small size. We propose for the first time to use a laser pulse-precursor, moving directly in front of the main pulse, to control the parameters of a self-injected and accelerated electron bunch. Previously, no one used a precursor to control the parameters of a self-injected and accelerated electron bunch.

The purpose of this paper is to show by the numerical simulation that some precursor-laser-pulse, moved directly before the main laser pulse, controls properties of the self-injected electron bunch and provides at certain conditions small energy spread and small size of self-injected and accelerated electron bunch.

## PARAMETERS OF SIMULATION

Fully relativistic particle–in–cell simulation was carried out by UMKA 2D3V code [30]. A computational domain (x, y) has a rectangular shape. $\lambda$ is the laser pulse wavelength. The computational time interval is $\tau = 0.05$, the number of particles per cell is 8 and the total number of particles is $15.96 \times 10^6$. The simulation of considered case was carried out up to 800 laser periods. The period of the laser pulse is $t_0 = 2\pi/\omega_0$. The s-polarized laser pulse enters into uniform plasma. In $y$ direction, the boundary conditions for particles, electric and magnetic fields are periodic. The critical plasma density $n_c = m_e \omega_0^2/(4\pi e^2)$. The laser pulse is defined with a "cos²" distribution in longitudinal direction. The pulse has a Gaussian profile in the transverse direction. The longitudinal and transverse dimensions of the laser pulse are selected to be shorter than the plasma wavelength. Directly in front of the main pulse a laser pulse-precursor moves. Pulse-precursor has a lower intensity and smaller radius compared to the intensity and radius of the main pulse (see Fig. 1). Pulse-precursor consists of two pulses of different intensities and radii, so that the intensity and radius of the whole pulse grow along the pulse from first its front. Full length at half maximum of the main laser pulse equals $\lambda$ and full width at half maximum equals $3\lambda$. $a_0 = eE_{z0}/(m_e c \omega_0) = 2.236$; $E_{z0}$ is the electric field amplitude. Full length at half maximum of the first pulse of the precursor equals $4\lambda$ and full width at half maximum equals $\lambda$, $a_{01} = eE_{z01}/(m_e c \omega_0) = 1$. Full length at half maximum of the second pulse of the precursor equals $2\lambda$ and full width at half maximum equals $2\lambda$, $a_{02} = eE_{z02}/(m_e c \omega_0) = 1.732$. Coordinates $x$ and $y$, time $t$, electric field amplitude $E_z$ and electron plasma density $n_0$ are given in dimensionless form in units of $\lambda$, $2\pi/\omega_0$, $m_e c \omega_0/(2\pi e)$, $m_e \omega_0^2/(16\pi^3 e^2)$.

## RESULTS OF SIMULATION

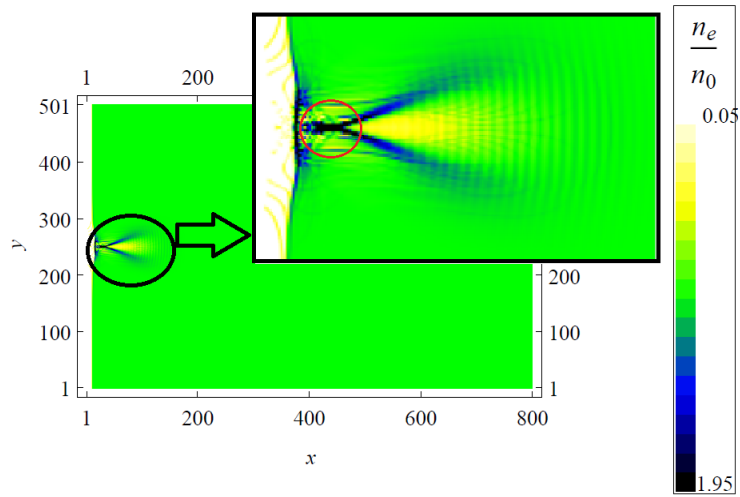

Fig. 1. Wake perturbation of plasma electron density $n_e$ excited by laser pulse near the boundary of injection

We consider the wakefield excitation by laser pulse, shaped on intensity and radius. The intensity and radius grow from the first front of the pulse and then abruptly break off. Let us show that one can control the wakefield using a precursor. Namely, stochastization of the wakefield is suppressed due to the adiabatic growth of intensity and radius of the pulse. Also one can control quality of the bunch. Namely, a point-kind bunch is formed.

From Fig.1 one can see that due to selected spatial distribution of intensity and radius of laser pulse near the boundary of injection the adiabatic structure, suitable for good self-injection, is formed.

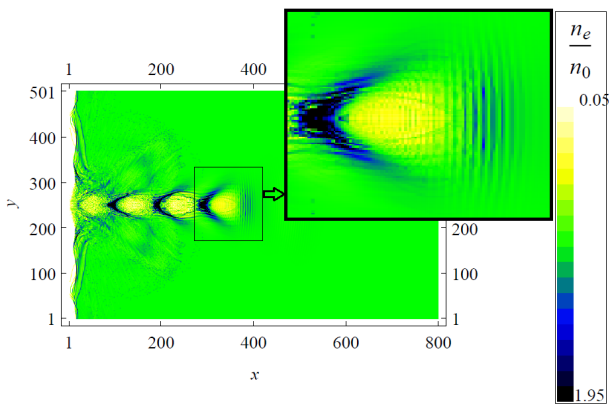

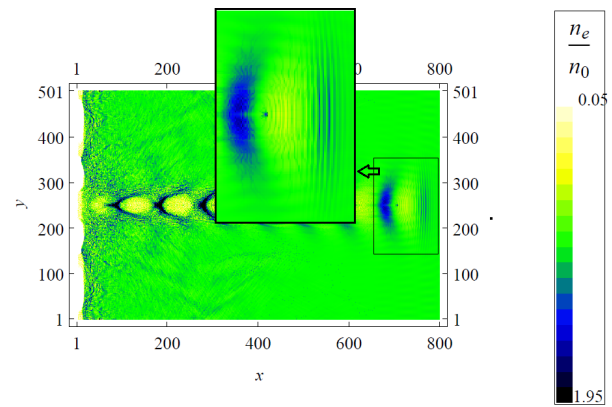

Fig. 2. Electron bunch self-injection at plasma wakefield excitation by laser pulse

Fig. 3. Self-injected electron bunch acceleration by wakefield bubble, excited in plasma by laser pulse

It is seen from Fig. 2 that short electron bunch is self-injected. Fig. 3 demonstrates that point-kind electron bunch is accelerated.

Stochastization of the wakefield is suppressed due to the adiabatic growth of intensity and radius of the laser pulse (Fig. 4).

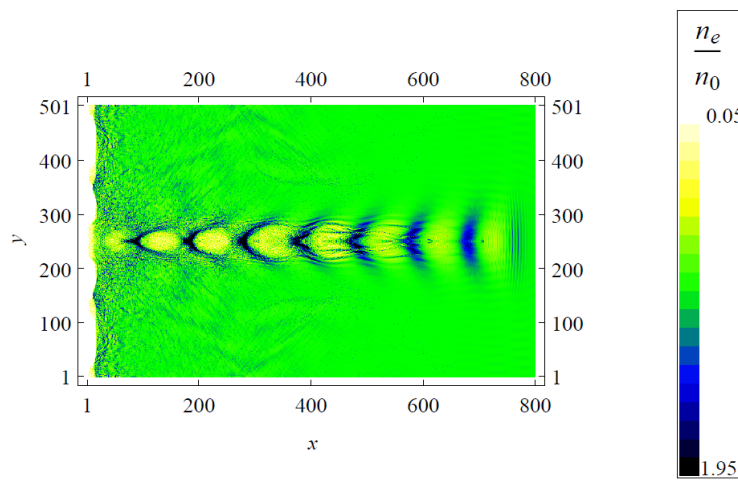

Fig. 4. Wakefield excitation and self-injected electron bunch acceleration in plasma by laser pulse

## CONCLUSION

Thus, the authors for the first time used a laser pulse-precursor, moving directly in front of the main pulse, to control the parameters of a self-injected and accelerated electron bunch. Previously, no one used a precursor to control the parameters of a self-injected and accelerated electron bunch. It has been shown by the numerical simulation that some precursor-laser-pulse, moved directly before the main laser pulse, controls properties of the self-injected electron bunch and provides at certain conditions small energy spread and small size of self-injected and accelerated electron bunch.

### ORCID IDs
**Vasyl Maslov** https://orcid.org/0000-0002-4370-7685, **Denys Bondar** https://orcid.org/0000-0002-7358-4305,
**Iryna Levchuk** https://orcid.org/0000-0003-0542-0410, **Ivan Onishchenko** http://orcid.org/0000-0002-8025-5825

### REFERENCES
[1]. A. Pukhov and J. Meyer-ter-Vehn, Apl. Phys. B., 7**4**, 355-361 (2002), doi: 10.1007/s003400200795.
[2]. K.V. Lotov, V.I. Maslov, I.N. Onishchenko and E.N. Svistun, Problems of Atomic Science and Technology, **6**, 114-116 (2008).
[3]. K.V. Lotov, V.I. Maslov, I.N. Onishchenko and E.N. Svistun, Plasma Phys. Cont. Fus. **52**, 065009 (2010), doi: 10.1088/0741-3335/52/6/065009.
[4]. K.V. Lotov, V.I. Maslov, I.N. Onishchenko and E.N. Svistun, Problems of Atomic Science and Technology. **3**, 159-163 (2012).
[5]. V.I. Maslov, I.N. Onishchenko and I.P. Yarovaya, Problems of Atomic Science and Technology. **1**, 134-136 (2013).
[6]. V.I. Maslov, I.N. Onishchenko and I.P. Yarovaya, East European Journal of Physics. **1**(2), 92-95 (2014).
[7]. I.P. Levchuk, V.I. Maslov and I.N. Onishchenko, Problems of Atomic Science and Technology. **4**, 120-123 (2015).

[8].  I.P. Levchuk, V.I. Maslov and I.N. Onishchenko, Problems of Atomic Science and Technology. **3**, 62-65 (2016).
[9].  K.V. Lotov, V.I. Maslov and I.N. Onishchenko and M.S. Vesnovskaya, Problems of Atomic Science and Technology. **4**, 12-16 (2010).
[10]. K.V. Lotov, V.I. Maslov, I.N. Onishchenko and M.S. Vesnovskaya, Problems of Atomic Science and Technology. **1**, 83-85 (2011).
[11]. V.A. Balakirev, I.N. Onishchenko and V.I. Maslov, Problems of Atomic Science and Technology. **3**, 92-95. (2011).
[12]. K.V. Lotov, V.I. Maslov and I.N. Onishchenko, Problems of Atomic Science and Technology. **4**, 85-89. (2010).
[13]. K.V. Lotov, V.I. Maslov and I.N. Onishchenko and I.P. Yarovaya, Problems of Atomic Science and Technology. **3**, 87-91 (2011).
[14]. V.I. Maslov, I.N. Onishchenko and I.P. Yarovaya, Problems of Atomic Science and Technology. **4**, 128-130 (2012).
[15]. V.I. Maslov, I.N. Onishchenko and I.P. Yarovaya, Problems of Atomic Science and Technology. **6**, 161 –163 (2012).
[16]. I.P. Levchuk, V.I. Maslov and I.N. Onishchenko, Problems of Atomic Science and Technology. **3**, 37 –41 (2015).
[17]. I.P. Levchuk, V.I. Maslov and I.N. Onishchenko, Problems of Atomic Science and Technology. **6**, 43 –46 (2017).
[18]. D.S. Bondar, I.P. Levchuk, V.I. Maslov and I.N. Onishchenko, East Eur. J. Phys. **5**(2), 72 –77 (2018).
[19]. K.V. Lotov, V.I. Maslov, I.N. Onishchenko and E.N. Svistun, Problems of Atomic Science and Technology. **2**, 122-124 (2010).
[20]. V. Lotov, V.I. Maslov, I.N. Onishchenko, E.N. Svistun and M.S. Vesnovskaya, Problems of Atomic Science and Technology. **6**, 114–116 ( 2010).
[21]. K.V. Lotov, V.I. Maslov and I.N. Onishchenko, Problems of Atomic Science and Technology. **6**, 103 –107 (2010).
[22]. K.V. Lotov, V.I. Maslov and I.N. Onishchenko, Problems of Atomic Science and Technology. **4**,73 –76 (2013).
[23]. W. P. Leemans, A. J. Gonsalves, H.-S. Mao, K. Nakamura, C. Benedetti, C. B. Schroeder, Cs. Tóth, J. Daniels, D. E. Mittelberger, S. S. Bulanov, J.-L. Vay, C. G. R. Geddes, and E. Esarey, Phys. Rev. Lett. **113**, 245002 (2014), doi: 10.1103/PhysRevLett.113.245002.
[24]. T. Tajima, Eur. Phys. J. Spec. Top. **223**, 1037–1044 (2014), doi: 10.1140/epjst/e2014-02154-6.
[25]. V.I. Maslov, O.M. Svystun, I.N. Onishchenko and V.I. Tkachenko, Nuclear Instruments and Methods in Physics Research A. **829**, 422–425 (2016), doi: 10.1016/j.nima.2016.04.018.
[26]. V.I. Maslov, O.M. Svystun, I.N. Onishchenko and A.M. Egorov, Problems of Atomic Science and Technology. **6**, 144-147 (2016).
[27]. D.S. Bondar, I.P. Levchuk, V.I. Maslov, S. Nikonova and I.N. Onishchenko, Problems of Atomic Science and Technology. **6**, 76-79 (2017).
[28]. D.S. Bondar, V.I. Maslov, I.P. Levchuk and I.N. Onishchenko, Problems of Atomic Science and Technology. **6**, 156 –159 (2018).
[29]. S. Hakimi, T. Nguyen, D. Farinella, C.K. Lau, H.-Yu. Wang, P. Taborek, F. Dollar and T. Tajima, Plas. Phys. **25**, 023112 (2018), doi: 10.1063/1.5016445.
[30]. G.I. Dudnikova, T.V. Liseykina, V.Yu. Bychenkov, Comp. Techn. **10**(1), 37 (2005).

**ПОЛІПШЕННЯ ВЛАСТИВОСТЕЙ САМОІНЖЕКТОВАНОГО І ПРИСКОРЕНОГО ЕЛЕКТРОННОГО ЗГУСТКУ ЛАЗЕРНИМ ІМПУЛЬСОМ В ПЛАЗМІ ВИКОРИСТАННЯМ ПЕРЕДВІСНИКА**

**В.І. Маслов[1,2], Д.С. Бондар[2], І.П. Левчук[1], І.М. Онищенко[1]**

*[1] Національний Науковий Центр «Харківський фізико-технічний інститут»*
*61108, Харків, вул. Академічна, 1*
*[2] Харківський національний університет імені В.Н. Каразіна*
*пл. Свободи 4, Харків, 61022, Україна*

Зараз прискорюючі поля в звичайних лінійних прискорювачах обмежені ~100 мВ/м. Прискорення в плазмі забезпечує прискорюючі поля, які на кілька порядків більше, ніж в звичайних прискорювачах. У зв'язку з успішним розвитком лазерних технологій, лазерно-плазмові прискорювачі зараз викликають великий інтерес. За минуле десятиліття успішні експерименти по лазерному прискоренню електронів в плазмі кільватерним полем підтвердили перспективність цього прискорення. Очевидно, що прискорюючі поля в лазерно-плазмових прискорювачах дозволяють зменшити розмір і знизити вартість прискорювачів. Інша важлива перевага лазерно-плазмових прискорювачів полягає в тому, що вони можуть створювати короткі електронні згустки великої енергії. Були продемонстровані електронні згустки з невеликим розкидом по енергії при взаємодії інтенсивних лазерних імпульсів з плазмою. Також була вивчена самоінжекція електронних згустків в кільватерній порожнині, яка генерується інтенсивним лазерним імпульсом у щільній плазмі. Завдяки нещодавно розвиненій компактній лазерній технології можна генерувати 100-ПВт лазерні одно-періоді фемтосекундні імпульси. За допомогою потужного фемтосекундного лазерного імпульсу можна генерувати когерентний рентгенівський імпульс. Професор Т.Таджіма запропонував використовувати ці когерентні рентгенівські імпульси для прискорення частинок. Коли такий рентгенівський імпульс інжектується в кристал, він взаємодіє з електронною плазмою металевої щільності і ідеально підходить для лазерного кільватерного прискорення. При числовому моделюванні авторів, виконаному на основі ідеї професора Т.Таджіми, при збудженні кільватерного поля рентгенівським лазерним імпульсом в електронній плазмі металевої густини було отримано прискорююче поле кілька ТВ/м. При лазерному прискоренні самоінжектованого згустку електронів кільватерним полем в плазмі важливо сформувати згусток з невеликим розкидом по енергії і невеликим розміром. У цій роботі числовим моделюванням показано, що певний імпульс-передвісник, що рухається перед основним лазерним імпульсом, контролює властивості самоінжектованого згустку і забезпечує за певних умов малий розкид по енергії і малий розмір самоінжектованого і прискореного електронного згустку.

**КЛЮЧОВІ СЛОВА:** короткий лазерний імпульс, кільватерне поле в плазмі, прискорення електронів, числове моделювання, самоінжекція електронних згустків

### УЛУЧШЕНИЕ СВОЙСТВ САМОИНЖЕКТИРОВАННОГО И УСКОРЕННОГО ЭЛЕКТРОННОГО СГУСТКА ЛАЗЕРНЫМ ИМПУЛЬСОМ В ПЛАЗМЕ ИСПОЛЬЗОВАНИЕМ ПРЕДВЕСТНИКА

**В.И. Маслов[1,2], Д.С. Бондарь[2] И.П. Левчук[1], И.Н. Онищенко[1]**
[1] *Национальный Научный Центр «Харьковский физико-технический институт»*
*61108, Харьков, ул. Академическая, 1*
[2] *Харьковский национальный университет имени В.Н. Каразина*
*пл. Свободы 4, Харьков, 61022, Украина*

В настоящее время ускоряющие поля в обычных линейных ускорителях ограничены ~100 МВ/м. Ускорение в плазме обеспечивает ускоряющие поля, которые на несколько порядков больше, чем в обычных ускорителях. В связи с бурным развитием лазерных технологий, лазерно-плазменные ускорители в настоящее время представляют большой интерес. За прошедшее десятилетие успешные эксперименты по лазерному ускорению электронов в плазме кильватерным полем подтвердили перспективность этого ускорения. Очевидно, что большие ускоряющие поля в лазерно-плазменных ускорителях позволяют уменьшить размер и снизить стоимость ускорителей. Другое важное преимущество лазерно-плазменных ускорителей заключается в том, что они могут создавать короткие электронные сгустки большой энергии. Были продемонстрированы электронные сгустки с небольшим разбросом по энергии при взаимодействии интенсивных лазерных импульсов с плазмой. Также была изучена самоинжекция электронных сгустков в кильватерной полости, генерируемой интенсивным лазерным импульсом в плотной плазме. Благодаря недавно развитой компактной лазерной технологии можно генерировать 100-ПВт лазерные одно-периодные фемтосекундные импульсы. С помощью мощного фемтосекундного лазерного импульса можно генерировать когерентный рентгеновский импульс. Профессор Т.Тадзима предложил использовать эти когерентные рентгеновские импульсы для ускорения частиц. Когда такой рентгеновский импульс инжектируется в кристалл, он взаимодействует с электронной плазмой металлической плотности и идеально подходит для лазерного кильватерного ускорения. При численном моделировании авторов, выполненном на основе идеи профессора Т.Тадзимы, при возбуждении кильватерного поля в электронной плазме металлической плотности было получено ускоряющее поле несколько ТВ/м. При лазерном ускорении самоинжектированного электронного сгустка кильватерным полем в плазме важно сформировать сгусток с небольшим разбросом по энергии и небольшим размером. В этой работе численным моделированием показано, что некоторый импульс-предвестник, движущийся перед основным лазерным импульсом, контролирует свойства самоинжектированного сгустка и обеспечивает при определенных условиях малый разброс по энергии и малый размер самоинжектированного и ускоренного электронного сгустка.
**КЛЮЧЕВЫЕ СЛОВА:** короткий лазерный импульс, кильватерное поле в плазме, ускорение электронов, численное моделирование, самоинжекция электронных сгустков